\documentclass[aps,prb,twocolumn, superscriptaddress]{revtex4}

\usepackage{graphicx}

\begin{document}


\title{Superconductivity in Pr$_{2}$Ba$_{4}$Cu$_{7}$O$_{15-\delta}$ 
with metallic double chains}

\author{M.Matsukawa} 
\email{matsukawa@iwate-u.ac.jp }
\affiliation{Department of Materials Science and Technology, Iwate University , Morioka 020-8551 , Japan }
\author{Yuh Yamada}
\affiliation{Department of Physics,Niigata University, Niigata 950-2181, Japan }
\author{M.Chiba}
\affiliation{Department of Materials Science and Technology, Iwate University , Morioka 020-8551 , Japan }
\author{H.Ogasawara}
\affiliation{Department of Materials Science and Technology, Iwate University , Morioka 020-8551 , Japan }
\author{T.Shibata}
\affiliation{Interdisciplinary Faculty of Engineering, Shimane University, Matsue  690-8504, Japan }
\author{A. Matsushita}
\author{Y.Takano}
\affiliation{National Institute for Materials Science, Tsukuba 305-0047 ,Japan}

\date{\today}

\begin{abstract}
We report superconductivity with $T_{c,onset}$=$\sim$16K in polycrystalline samples of \\
Pr$_{2}$Ba$_{4}$Cu$_{7}$O$_{15-\delta}$ possessing metallic double chains. A reduction treatment on as-sintered samples causes not only the enhanced metallic 
conduction but also the appearance of superconductivity. This finding is strongly contrast with the oxygen removal effect of superconducting Y$_{2}$Ba$_{4}$Cu$_{7}$O$_{15-\delta}$. 

\end{abstract}


\maketitle

\begin{figure}[h]
\includegraphics[width=8cm]{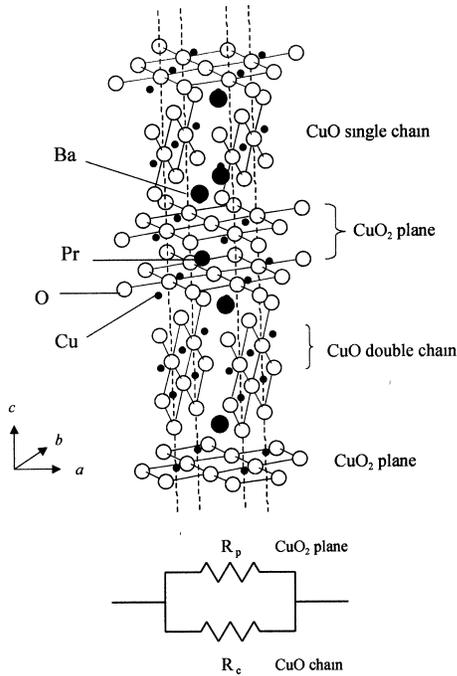}%
\caption{\label{St} Schematic view of the 
Pr$_{2}$Ba$_{4}$Cu$_{7}$O$_{15-\delta}$ structure with a repetition of the corner-sharing CuO single chain 
and the egde-sharing CuO double chains. 
Also shown is the direction of the orthorhombic principal axes. The inset indicates models of parallel resistors 
along the $b$ direction
,where R$_{p}$ and R$_{c}$denote resistances of CuO$_{2}$ planes and CuO chains, respectively.  }
\end{figure}%

Since the discovery of high-$T_{c}$ copper-oxide superconductors,  extensive studies 
on strongly electron correlated system  have been in progress on the basis of physical properties 
of two-dimensional (2D) CuO$_{2}$ planes.  Moreover, from the viewpoint of low-dimensional physics, 
particular attention is paid to the physical role of one-dimensional (1D) CuO chains included 
in some families of high-$T_{c}$ copper oxides such as Y-based superconductors with the transition 
temperature $T_{c}$=$\sim$ 92K .  
It  is well known that the Pr-substitution for  Y-sites in YBa$_{2}$Cu$_{3}$O$_{7-\delta}$ (Y123/7-$\delta$) 
and  YBa$_{2}$Cu$_{4}$O$_{8}$ (Y124/8) compounds dramatically suppresses $T_{c}$  and 
superconductivity in CuO$_{2}$ planes disappears beyond the critical value of  Pr , $x_{c}$=0.5 and 0.8, 
respectively \cite{SO87,HO98}. Such a suppression effect due to Pr-substitution on superconductivity 
has been explained in terms of the hybridization model with respect to 
Pr-4$f$ and O-2$p$ orbitals \cite{FE93}.Y124/8 compound with double chains is thermally stable up to 
800 $^\circ$C 
,while in Y123/7-$\delta$ oxygen deficiencies are easily introduced at lower annealing temperatures 
\cite{JO87}.  
Intermediate between PrBa$_{2}$Cu$_{3}$O$_{7-\delta}$ 
(Pr123/7-$\delta$) with single chains and PrBa$_{2}$Cu$_{4}$O$_{8}$
(Pr124/8) with double chains   is  the Pr$_{2}$Ba$_{4}$Cu$_{7}$O$_{15-\delta}$ (Pr247/15-$\delta$) compound 
with an alternative repetition of the single and double chains along the $c$-axis,
isostructural with superconducting Y$_{2}$Ba$_{4}$Cu$_{7}$O$_{15-\delta}$ (Y247/15-$\delta$)  \cite{BO88,YA94}. 
The crystal structure of  orthorhombic Pr247 is schematically shown in Fig. 1 (space group $Ammm$).
In Pr247, it is possible to examine physical properties of metallic double chains 
varying oxygen contents along single chains \cite{YA94,MA95}.

In this paper,  we report  superconductivity in 
Pr$_{2}$Ba$_{4}$Cu$_{7}$O$_{15-\delta}$ compound possessing metallic double chains.
Polycrystalline Pr$_{2}$Ba$_{4}$Cu$_{7}$O$_{15-\delta}$ compound  was  synthesized  using a powder sintering method 
under high pressure oxygen \cite{YA94,HO98}.  The Pr$_{6}$O$_{11}$, BaO$_{2}$ and CuO powders with high purity 
were mixed to the stoichiometric composition and then were pressed into a cylindrical pellet. 
The pellet was calcined at 850-900 $^\circ$C in air and then was  sintered at 975 $^\circ$C for 18 hours 
with P(O$_{2}$)=5 atm. As-sintered samples were annealed in argon gas at 650 $^\circ$C. 
The resistivity measurement was performed by a conventional four-probe technique. 
The magnetization  measurement  was carried out  down to 2K under zero-field cooling (ZFC) and field cooling(FC)conditions 
using a SQUID magnetometer. 


\begin{figure}
\includegraphics[width=8cm]{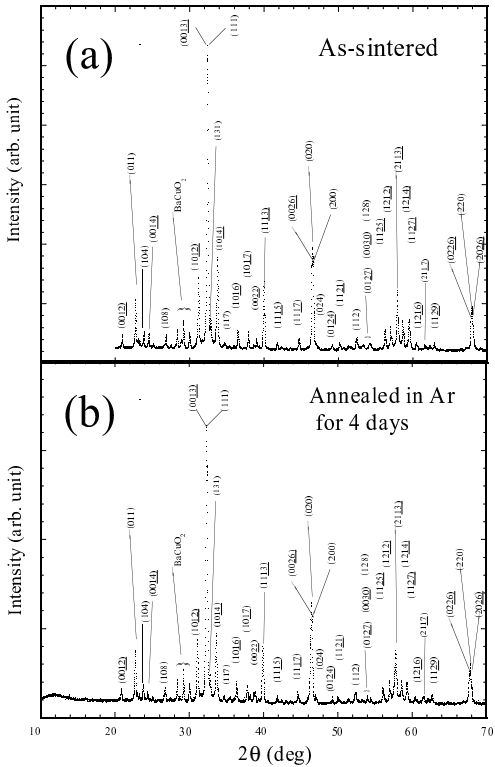}%
\caption{\label{Res}  X-ray diffraction patterns of Pr$_{2}$Ba$_{4}$Cu$_{7}$O$_{15-\delta}$ 
compounds at room temperature  : (a) as-sintered and (b) annealed in Ar for 4 days.
The indices represent reflections of the Pr247 phase.} 
\end{figure}%

 ~~~Figure 2 shows the X-ray diffraction patterns for as-sintered and Ar reduced 
Pr247/15-$\delta$ at room temperature.
All diffraction lines were indexed in terms of the Pr247 majority phase,  except for  a small amount of the BaCuO$_{2}$ phase. \\
~~~Here, we give some comments on impurity phases in Pr247.
From our previous studies on synthesis of Pr247 compounds, we have pointed out 
possible impurity phases such as  PrBaO$_{3}$ , BaCuO$_{2}$ and Pr124/8 
, except for the starting materials \cite{YA94,HO98}.  X-ray diffraction patterns on 
both as-sintered and reduced Pr247 reveal@a small amount 
of the BaCuO$_{2}$  impurity phase although the precipitation of  PrBaO$_{3}$  and 
Pr124/8 is strongly suppressed in this sintering process.  
 It is truly expected that in the present samples, the PrBaO$_{3}$  and Pr124/8 phases 
may be included as a very small volume fraction hardly 
detectable by x-ray diffraction. However, the lower-$T$ heat treatment  never causes  
superconductivity for the PrBaO$_{3}$ , BaCuO$_{2}$  
and Pr124/8  phases. 
The oxide materials of  PrBaO$_{3}$  and BaCuO$_{2}$  are highly insulator. Moreover, the 
Pr124/8 phase is thermally stable under lower temperature annealing 
and the resistivity data of  metallic Pr 124 single crystals exhibit no 
superconductivity down to 2K\cite{HO00}. Therefore,  we believe  that even if there 
exist minor impurity phases hardly observed by x-ray probe, low-$T$ reduction process induces 
no superconductivity for the impurity phases in Pr247. \\
~~~The lattice parameters of the as-sintered sample are estimated from the least-squared fitting to the X-ray diffraction data to be 
$a$=3.8880(3),  $b$=3.9037(3) and $c$=50.660(4) \AA\,  respectively. The oxygen deficiency $\delta$ of as-sintered Pr247 is taken as almost zero because pure Y247prepared under the same sintering condition as Pr247, exhibits a sharp superconducting  transition with $T_{c}$=$\sim$89K \cite{HO98}.  In fact, Tallon $et \, al.$, reported that the resistivity of Y247/15 shows a superconducting transition near $T_{c}$=$\sim$92K\cite{TA90}.
For comparison, the unit length of $c$-axis in Pr247/15  [=2$\times$(Pr123/7 
unit)+(Pr124/8 unit)]  is estimated to be 50.77 \AA\ from
$c$=11.73 \AA\ for Pr123/7 and $c$=27.308 \AA\ for Pr124/8 \cite{LO90,YA94}. 
This estimation is almost comparable with
the $c$-axis lattice constant of Pr247/15. For the sample reduced in Ar for 4 days, 
the lattice parameters are determined to
be $a$=3.8923(5),  $b$=3.9020(3) and $c$=50.814(6) \AA\,  respectively. The oxygen 
deficiency $\delta$ of  Pr247 due to Ar annealing is estimated to be $\delta$=$\sim$0.5 from the thermogravimetry and results in a substantial
elongation of the $c$-axis length up to $\sim$0.3\%,  keeping its orthorhombic 
structure.  These results are similar with a variation in the lattice
parameters of Y247 as a function of oxygen deficiency.  Irizawa $et al.$ , reported that 
the $c$-axis length
of Y247/15-$\delta$ reaches an increase by 0.36\% over the range of oxygen deficiency 
from $\delta$=0.05 to 0.6
\cite{IR02}. \\

\begin{figure}[ht]
\begin{center}
\includegraphics[width=8cm]{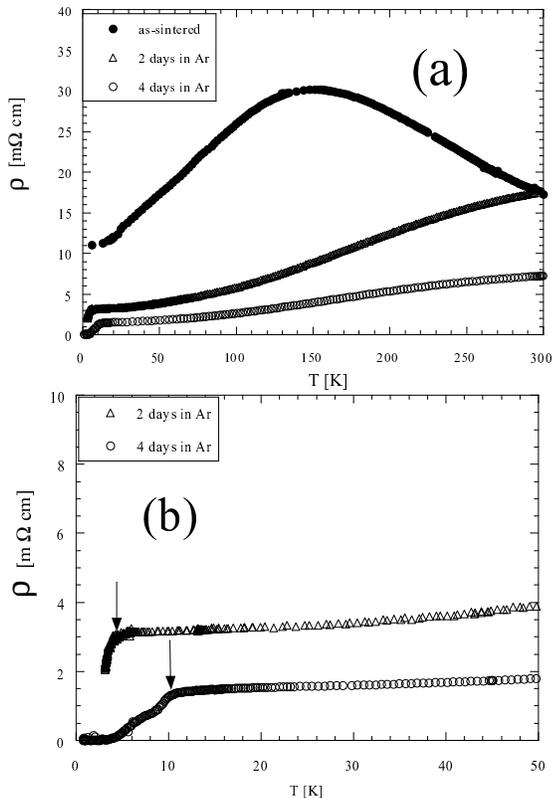}%
\end{center}
\caption{The temperature variation of electrical resistivity in  Pr247 , 
(a)over the whole range of temperature up to 300K  ,and (b) below 50 K. 
The arrows denote the value of $T_{c,onset}$. }
\end{figure}%

\begin{figure}
\includegraphics[width=8cm]{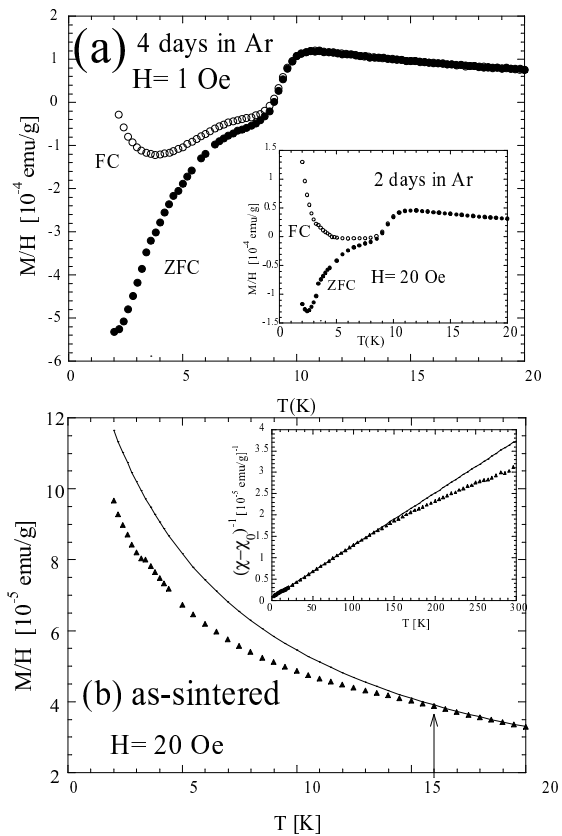}%
\caption{\label{MT}(a) The magnetic susceptibility($M/H$) of the 4-days reduced sample  in Pr247 
 as a function of temperature under  zero field cooling (ZFC) and field cooling (FC) conditions
 at 1 Oe.In the inset, the $M/H$ data of the 2-days one are also shown. 
(b)The susceptibility data of the as-sintered sample in Pr247 . A solid curve represents 
a least-squared fitting according to the Curie-Weiss law taken into account 
the $T$ independent component$\chi_{0}$.The arrow denotes the antiferromagnetic ordering temperature of Pr ions.
The inset of Fig.4(b)indicates ($\chi-\chi_{0}$)$^{-1}$ versus $T$ plots up to 300K.}
\end{figure}%
\begin{figure}[ht]
\begin{center}
\includegraphics[width=8cm]{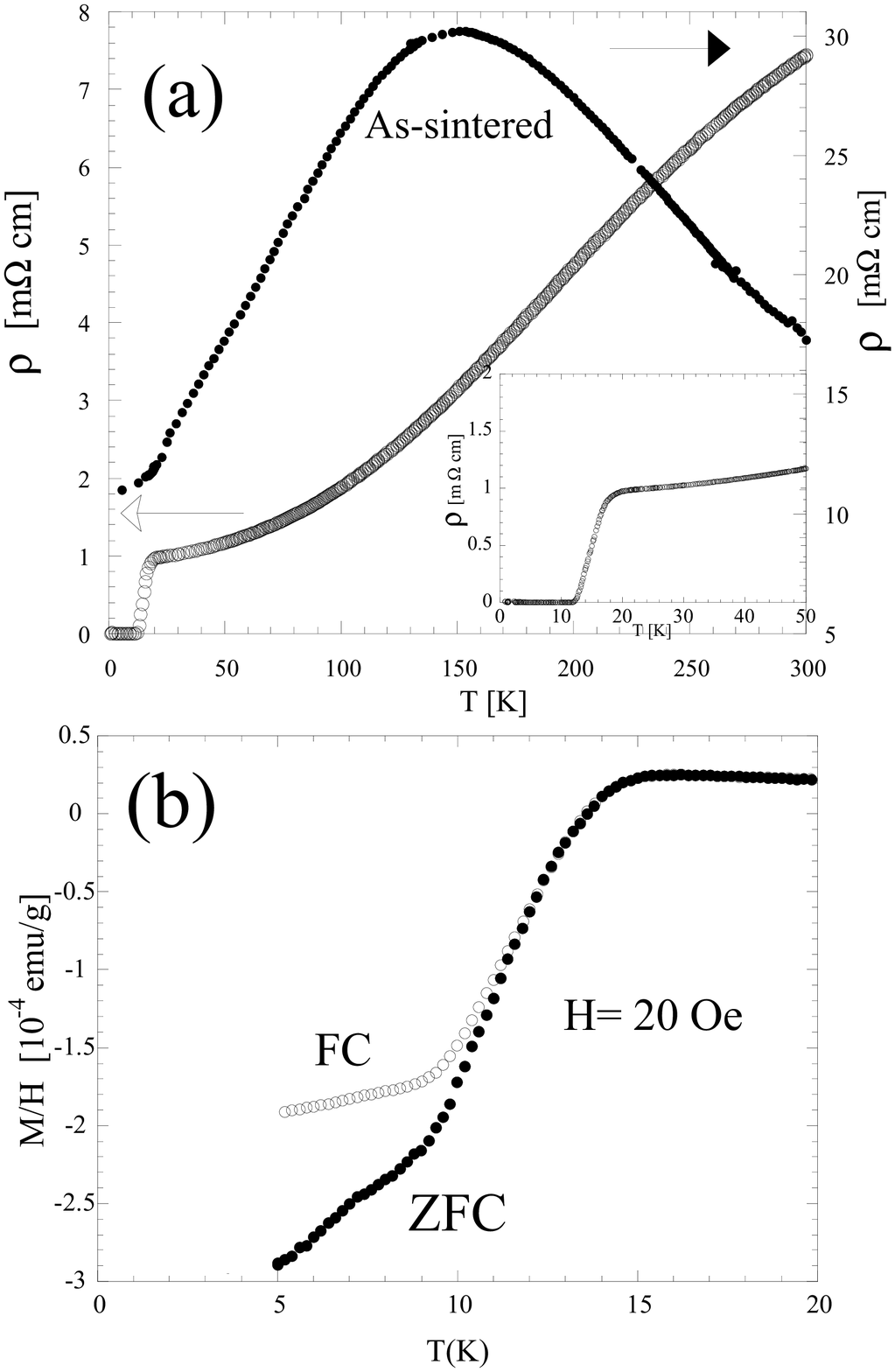}%
\end{center}
\caption{(a)The resistivity and (b) magnetization data for Pr247 samples annealed in a
vacuum (at 400$^\circ$C for 24 hours) as a function of temperature. In the inset of Fig.5 (a), its magnified data below 50 K are shown. A low-temperature vacuum reduction enhances the value of $T_{c,onset}$=$\sim$16K.  
}
\end{figure}%

~~~Figure 3(a) shows the temperature variation of electrical resistivity  in sintered 
Pr247/15-$\delta$. The value of $\rho$ of  the as-sintered sample exhibits a semiconducting behavior at high $T$, it reaches a broad maximum at  $T_{m}$= $\sim$150K and  gradually decreases down to 4K.  This finding has been well explained  on the basis of  the parallel circuit model (the inset of Fig.1) composed  of
 both the semiconducting CuO$_{2}$ planes  and metallic CuO double chains, as first 
proposed by Bucher $et al$,  in YBCO system \cite{MA95,BU90}. 
A previous study on both sintered Pr 124 and Pr247  implicated  that  the 
negative temperature dependence of the resistivity above $T_{m}$ is ascribed to the 
CuO$_{2}$ plane conduction,  while the positive $T$-dependence of the 
resistivity below  $T_{m}$ to the CuO double-chain conduction \cite{MA95}.   
 On the other hand, the $T$ dependence of  $\rho$ of the reduced samples follows a 
strongly metallic conduction  up to 300K.
For the 4-days reduced sample,  the superconducting transition  evidently appears 
around 10 K and the zero-resistivity temperature $T_{c,zero}$ is determined to be 3.8K from
 low-$T $ resistivity data. 
The value of $\rho$ of the superconducting sample is proportional to $T^2$  below 
150K,  which is the same temperature dependence of  $\rho$ along the $b$-axis of Pr 124 
single crystal  \cite{HO00}. 
Accordingly, the oxygen defects in Pr247 system cause not only the enhanced 
metallic state but also the appearance of superconductivity,  in strongly contrast with the oxygen removal effect of Y247 on its resistivity \cite{IR02}. \\
~~~Next, the temperature dependence of magnetic susceptibility ($\chi$) in Pr247 is 
shown in Fig.4. 
 For the 4-days reduced sample, the value of  $\chi$  gradually increases  with 
decreasing $T$,
 then  exhibits a hump  near 10K and finally a clear diamagnetic signal is detected  
below $\sim$9 K on both ZFC  and FC data. The superconducting volume fraction is
 estimated to be about 4\% at 2K  from the ZFC data.  Moreover, for 2-days reduced one, 
a sharp drop in the $\chi$  data  is also observed below 10K.  
On the other hand,  the value of $\chi$ for the as-sintered one  shows  a Curie-Weiss 
like  behavior  and  a slight anomaly is observed  at $\sim$3K indicating the formation 
of superconducting state.  
The Curie-Weiss plot suggests that the N\'{e}el temperature of Pr spins $T_{N}$ 
is $\sim$ 16K almost the same as $T_{N}$(=17K)in Pr124 \cite{HO98}.
Furthermore, the ($\chi-\chi_{0}$)$^{-1}$ data in the inset of Fig.4(b)
show a nonlinear deviation near $\sim$ 160 K associated with the N\'{e}el ordering 
of Cu spins, which is not far from  $T_{N}$(=200 K)in Pr124 \cite{LI99}. 
This finding is consistent with the evidence for an  antiferromagnetic order of the planar Cu spins in the NQR experiment \cite{FU03}. \\
~~~Recently,  we have found out that a low-$T$ reduction treatment in a vacuum causes a higher superconductive sample of Pr247, as shown in Fig.5, through our seeking for more suitable annealing conditions.  A  sharply superconducting transition is observed around 16K and the zero resistivity state is completely achieved at 13K, in spite of  a small fraction of superconductivity (4\% at 2K, the data of $M(T)$ below 5K are not shown here) .  However, a longer heat treatment on as-sintered samples in a vacuum than 24 hours do not  improve their superconducting property accompanied by a saturation in the $c$-axis elongation, where the $c$ axis length expands up to 50.93 \AA\ under annealing at 400 $^\circ$C for 24 hours.    It should be noted that  superconductive samples are obtained reproducibly  under the similar annealing procedure from starting materials.  We wonder why the $\rho$=0 state is realized in the bulk sample of Pr247, in spite of a small volume fraction of superconductivity estimated from low-$T$ magnetization data.  In granular superconductive system, a percolative model predicts a normal to superconduting transition, in other words, the $\rho$=0 state through the formation of superconducting paths beyond a threshold of superconductive volume fraction, where  its critical value is about 15\% in 3D case.   
For example, in a Au-Y123 granular system,  the zero resistivity state disappeared for the samples investigated below the net superconducting volume fraction of  20 \%,  in spite that in their samples a clear diamagnetic signal with $T_{c}$=90K  was detected \cite{XI88}.  In a MgO-Bi2212 composite system, the non zero-resistance  was also reported near the volume fraction of 12\% while its superconductive grains of Bi2212 matrix exhibited no drop of  $T_{c}$=70K from $M(T)$ data \cite{LI93}.  
Polycrystalline samples of Pr247 prepared by a sintering process are composed of  numerous grains with their grain size of  several micron meters.  Following the percolation model, several percents of superconductive volume fraction causes at most a gradual  drop in resistivity but never realize the zero resistive state for polycrystalline samples.
On the contrary,  the $\rho$=0 state is surely realized in reduced Pr247 accompanied by the enhanced  metallic state.  We expect that a small fraction within individual grains of Pr247 contributes superconducting paths, forming a zero resistive network.  This finding suggests that the observed superconductivity is closely related to a bulk property of Pr247.
   
~~~Here, we would like to comment on superconductivity with $T_{c}$=85K  in a Pr123 
single crystal 
reported by Zou $et\, al$ \cite{ZO98}.  Surely, the origin of  bulk superconductivity in 
Pr123 has not been made clear,  but several authors claimed that the superconductivity originates from the inhomogeneities in the chemical composition such as the Ba-rich Pr123 \cite{NA99}.  The  partial  substitution  of  Ba  for  Pr -site  gives  rise  to a strong  
suppression  of  the  hybridization  between  Pr-$4f$  and  O-$2p$ orbits,  
resulting  in  doping  mobile carriers into CuO$_{2}$  planes and the appearance of superconductivity.  This issue seems to depend on the criterion whether  the orbital 
hybridization leading to carrier localization is retained or not.  The negative 
$T$-dependence of  resistivity for
 as-sintered Pr247 (Fig.3(a)) demonstrates a clear contribution from the insulating 
CuO$_{2}$ planes as well as the insulating Pr123. 
 Moreover,  Matsushita  $et\, al$  showed  that  the value of  $\rho$  in  
as-sintered Pr247 gradually rises 
at high temperatures with applied pressure, indicating the enhancement of  the 
hybridization state between Pr and O orbits
\cite{MA95}.  It is easily understood that  a reduction process does not  alter  the  
electronic state of  
the Pr-O hybridization  because a lower temperature annealing in Ar hardly cause the 
Ba substitution for Pr sites 
in the Pr247 composition. In fact, a previous work on stoichiometric 
Pr123/7-$\delta$ reported that its resistivity 
increases by more than three orders of magnitude  as  the  oxygen content 
7-$\delta$ is reduced 
from ~6.93 to 6.46 \cite{LO90}.
In this study, the $c$-axis elongation due to oxygen deficiency  gives no changes  
on  the  Pr-O  hybridization state
in Pr123 system.  In a similar way, it is expected that there also exists the 
hybridization in  the  reduced Pr247 
accompanied by the $c$-axis elongation due to oxygen removal.
In the view point of sample preparation, we have to point out outstanding differences 
between Pr123 and Pr247.  
As-grown crystal of Pr123 was annealed in flowing
pure oxygen gas for over 4 days under two steps of annealing temperatures in order 
to realize superconductivity.
Zou et al., reported that the $c$-axis lattice constant  of the annealed crystal is 
shorter than that of the as-grown one.
Therefore,  it  is  difficult  to  identify  the  observed  superconductivity  
in reduced Pr247 with 
the bulk superconductivity in oxidized Pr123 crystal. 

Moreover, Hall coefficient measurement \cite{MA04} on reduced Pr247 reveals that  R$_{H}$($T$) clearly shows its negative sign below 100K, which is probably related to the enhanced metallic state due to oxygen removal.
A systematic study on the pressure effect on  both transport and magnetic properties in the superconducting Pr247 is in progress \cite{YA03}.
A microscopic research such as the NMR will be also carried out,  in order to specify 
the superconducting regions.  \\
~~~In summary, we have reported superconductivity in polycrystalline samples of 
Pr$_{2}$Ba$_{4}$Cu$_{7}$O$_{15-\delta}$ possessing metallic double 
chains. Both resistivity and magnetization measurements exhibit the superconducting 
transition temperature $T_{c,onset}$=$\sim$ 16K and the superconducting volume fraction is estimated to be about 4\% at 2K.
The reduction treatment  causes not only the enhanced metallic conduction but also
the realization for superconductivity, accompanied by the c-axis elongation due to 
oxygen deficiency. 
%


%


\end{document}